\documentclass[a4paper,11pt]{article}
\usepackage{pos}

\title{Spatial and Transverse structure of Heavy B- and D-mesons}

\author*[a]{Satyajit Puhan}
\author[a]{Harleen Dahiya}

\affiliation[a]{Dr. B R Ambedkar National Institute of Technology,\\
  Jalandhar, Punjab, India}

\emailAdd{puhansatyajit@gmail.com}
\emailAdd{dahiyah@nitj.ac.in}

\abstract{ We have investigated the unpolarized valence quark generalized parton distribution functions (GPDs) and  parton distribution functions (PDFs) for heavy spin-$0$, $B$- and $D$-mesons in the light-front quark model (LFQM). PDFs have been extracted from unpolarized $f_1(x,\textbf{k}^2_{\perp})$ transverse momentum-dependent parton distribution functions (TMDs). We have solved the quark-quark correlation function to have an unpolarized $H(x,\zeta,t)$ GPD. The unpolarized GPDs at zero skewness ($\zeta=0$) lead to describing the electromagnetic form factors (EMFFs) ($F_M(t)$) and gravitational form factors (GFFs)($A_M(t)$) of the mesons.}

\FullConference{
16th International Conference on Heavy Quarks and Leptons (HQL2023)\\
28 November-2 December 2023\\
TIFR, Mumbai, Maharashtra, India\\}

\begin{document}
\maketitle

\section{Introduction} Distribution functions (DFs) \cite{Diehl:2015uka} have been widely used to describe the multidimensional structure of the elementary particles. In this context, TMDs \cite{Puhan:2023ekt} and GPDs \cite{Meissner:2008ay} play an important role to understand the distributions of constituents inside the mesons. The GPDs provides the information about spatial structure ($\zeta,t$) and longitudinal momentum fraction ($x$), while the TMDs carry the information about transverse momenta ($\textbf{k}_{\perp}$) of the quark along with $x$. GPDs are accessed through the deeply virtual compton scattering (DVCS) and deeply virtual meson production (DVMP) processes. TMDs can be extracted from the deep inelastic scattering (DIS) processes, Drell-Yan processes and semi-inclusive deep inelastic scattering (SIDIS) processes. In this present work, we have discussed the unpolarized TMD and GPD for heavy mesons using the light-front quark model.

\section{Light-Front Quark Model}
The light-front (LF) minimal Fock-state representation for mesons with momentum $P$ is
\begin{eqnarray}
|M(P, S_Z)\rangle &=& \sum_{\lambda_i,\lambda_j}\int
\frac{\mathrm{d} x \mathrm{d}^2
        \mathbf{k}_{\perp}}{\sqrt{x(1-x)}16\pi^3}
           \Psi_{S_Z}(x,\mathbf{k}_{\perp},\lambda_i,\lambda_j)|x,\mathbf{k}_{\perp},
        \lambda_i,\lambda_j \rangle
        .
        \label{meson}
\end{eqnarray}
Here $x=\frac{k^+}{P^+}$ and $\mathbf{k}_{\perp}$ are the longitudinal momentum fraction and transverse momentum of the active quark respectively. $\Psi_{S_Z}(x,\mathbf{k}_{\perp},\lambda_i,\lambda_j)$ is the LF meson wave function with different spin and helicity projections $\lambda$. It can be expressed as 
\begin{eqnarray}
\Psi_{S_z}(x,\textbf{k}_\perp, \lambda_i, \lambda_j)= J_{S_z}(x,\textbf{k}_\perp, \lambda_i, \lambda_j) \psi^{M}(x, \textbf{k}_\perp).\
\label{space}
\end{eqnarray}
Here $J_{S_z}(x,\textbf{k}_\perp, \lambda_i, \lambda_j)$ and $\psi^{M}(x, \textbf{k}_\perp)$ are the spin and momentum space wave functions of the mesons respectively. 
The momentum space wave function in Eq. \ref{space} can be expressed using  Brodsky-Huang-Lepage
(BHL) \cite{Qian:2008px} as 
\begin{eqnarray}
\psi^M(x,\textbf{k}_\perp)= A \ {\rm exp} \Bigg[-\frac{\frac{\textbf{k}^2_\perp+m_q^2}{x}+\frac{\textbf{k}^2_\perp+m^2_{\bar q}}{1-x}}{8 \beta^2}
-\frac{(m_q^2-m_{\bar q}^2)^2}{8 \beta^2 \bigg(\frac{\textbf{k}^2_\perp+m_q^2}{x}+\frac{\textbf{k}^2_\perp+m_{\bar q}^2}{1-x}\bigg)}\Bigg],
\label{bhl-k}
\end{eqnarray}
where $m_{q (\bar q)}$ is the masses of quark (anti-quark) of the meson. $A$ and $\beta$ are the normalization constant and harmonic scale parameter of the respective mesons respectively. The input parameters have been taken from Ref. \cite{Arifi:2022pal}.

\par The spin wave function for pseudoscalar mesons ($S_z=0$) with different helicites is expressed as \cite{Qian:2008px}
\begin{equation}
\left\{
  \begin{array}{lll}
    J^{(S_z=0)}_P(x,\mathbf{k}_\perp,\uparrow,\uparrow)&=&\frac{1}{\sqrt{2}}\omega^{-1}(-k^L)(M+m_q+m_{\bar q}),\\
    J^{(S_z=0)}_P(x,\mathbf{k}_\perp,\uparrow,\downarrow)&=&\frac{1}{\sqrt{2}}\omega^{-1}((1-x)m_q+x m_{\bar q})(M+m_q+m_{\bar q}),\\
    J^{(S_z=0)}_P(x,\mathbf{k}_\perp,\downarrow,\uparrow)&=&\frac{1}{\sqrt{2}}\omega^{-1}(-(1-x)m_q-x m_{\bar q})(M+m_q+m_{\bar q}),\\
    J^{(S_z=0)}_P(x,\mathbf{k}_\perp,\downarrow,\downarrow)&=&\frac{1}{\sqrt{2}}\omega^{-1}(-k^{R})(M+m_q+m_{\bar q}),
  \end{array}
\right.
\end{equation}
with
$\omega=(M+m_q+m_{\bar q})\sqrt{x(1-x)[M^2-(m_q-m_{\bar q})^2]}$ and $k^{L(R)}=k_x \pm k_y$. $M$ is the mass of hadrons expressed as 
\begin{eqnarray} 
   M^2 &=& \frac{m^2_{q}+\textbf{k}^2_{\perp}}{x} + \frac{m^2_{\bar q}+\textbf{k}^2_{\perp}}{1-x}.
\end{eqnarray}

\section{Transverse momentum-dependent parton distributions}
In case of the pseudo-scalar mesons, the leading twist unpolarized TMD is expressed through the quark-quark correlation function is defined as \cite{Meissner:2008ay}
    \begin{equation}\label{coor}
      f^{q[\gamma^+]}_1(x,\textbf{k}^2_\perp)=\frac{1}{2}\int\frac{\mathrm{d}z^-\mathrm{d}^2\vec{z}_\perp}{2(2\pi)^3}e^{i \textbf{k}\cdot z} \langle M|\bar\Psi(0)\gamma^{+} \mathcal{W}(0,z)\Psi(z)|M \rangle|_{z^{+}=0}.
    \end{equation}
       The unpolarized $f^{q}_{1}(x,\textbf{k}^{2}_{\perp)}$ TMD in the form of wave function is found to be 
       \begin{eqnarray}
f_1^{q}(x,\textbf{k}^2_\perp)&=&\frac{1}{16 \pi^3} \big[\mid{ \psi(x,\textbf{k}_\perp, \uparrow, \uparrow )}\mid^2 +\mid \psi (x,\textbf{k}_\perp, \downarrow, \downarrow )\mid^2 + \mid \psi  (x,\textbf{k}_\perp, \downarrow, \uparrow )\mid^2 \nonumber\\
&& + \mid \psi (x,\textbf{k}_\perp, \uparrow, \downarrow)\mid^2\big].
\end{eqnarray}
The unpolarized PDF is given by 
\begin{eqnarray}
    f^{q}(x)= \int_{0}^{\infty} d\textbf{k}^2_{\perp} f_1^{q}(x,\textbf{k}^2_\perp).
\end{eqnarray}
All the PDFs obey the sum rule 
\begin{equation}
  \int_{0}^{1} d x f^q(x) =1\,.  \end{equation}

  \begin{figure}
      \centering
      \includegraphics[scale=0.3]{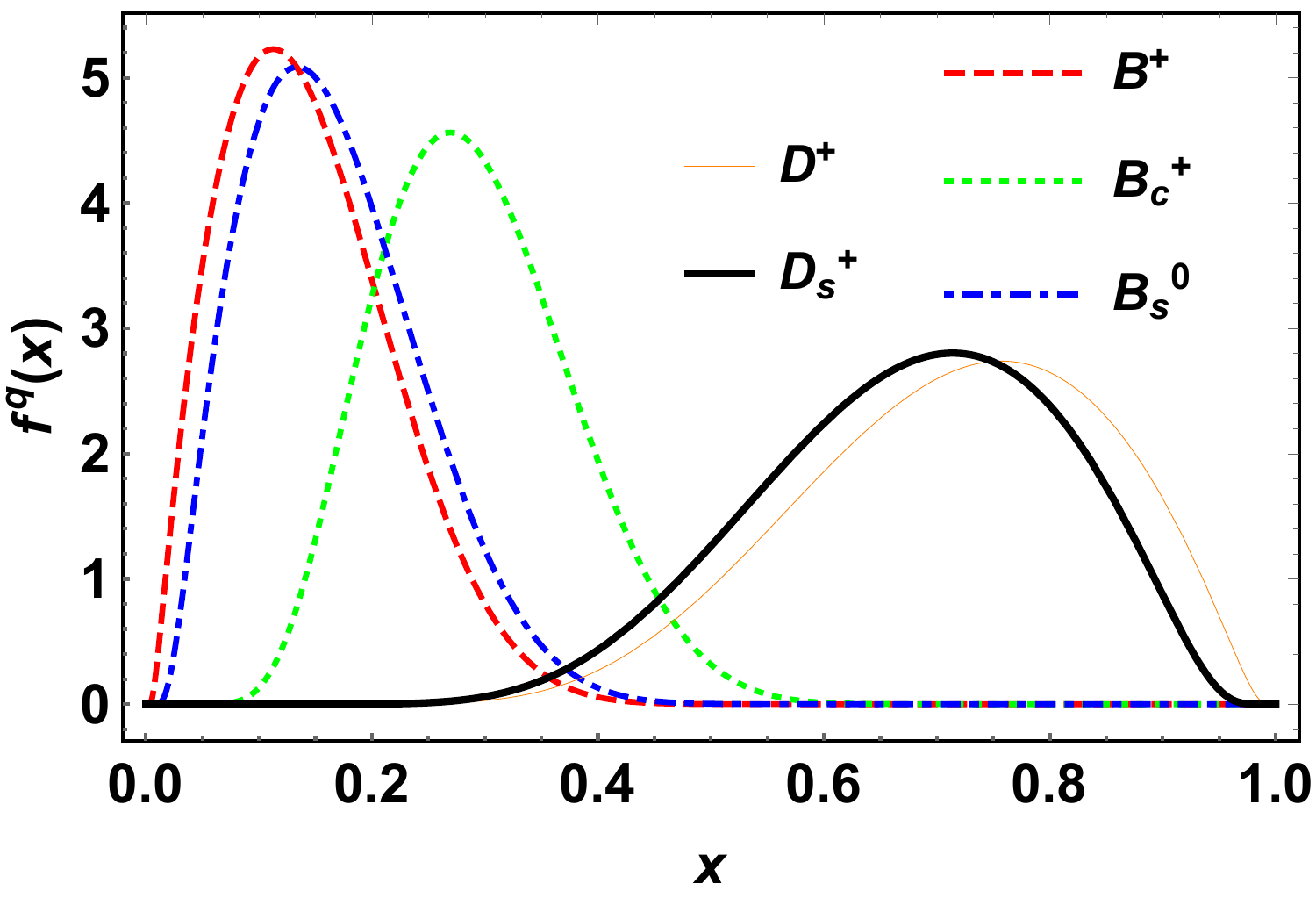}
      \caption{Comparison of PDFs for $B$- and $D$-mesons.}
      \label{pdf}
  \end{figure}
  \section{Generalized Parton Distributions}
  The unpolarized GPD correlation function at leading twist with initial momentum $P''$ and final momentum $P'$ is expressed as \cite{Meissner:2008ay} \begin{eqnarray}
H(x,\zeta,t)&=& \int \frac{dz^-}{4 \pi} e^{i k^+ z^-/2} \langle{M(P'')}|\bar{\Psi}(0)\gamma^+ \Psi(z)|M(P') \rangle \vert_{z^+=\textbf{z}_\perp=0}.
\end{eqnarray}
At $\zeta=0$, $H(x,0,t)$ is found to be 
\begin{eqnarray}
    H(x,0,t)=\int \frac{d^2\textbf{k}_\perp}{16 \pi^3} \big[\psi^*(x,\textbf{k}''_\perp, \uparrow, \uparrow )\psi(x,\textbf{k}'_\perp, \uparrow, \uparrow)+\psi^*(x,\textbf{k}''_\perp, \uparrow, \downarrow) \psi(x,\textbf{k}'_\perp, \uparrow, \downarrow )\nonumber\\
   +\psi^*(x,\textbf{k}''_\perp, \downarrow, \uparrow )\psi(x,\textbf{k}'_\perp, \downarrow, \uparrow )+\psi^*(x,\textbf{k}''_\perp, \downarrow, \downarrow )\psi(x,\textbf{k}'_\perp, \downarrow, \downarrow )\big],
\end{eqnarray}
with quark final and initial momenta,
\begin{eqnarray}
    \textbf{k}''_\perp=\textbf{ k}_\perp -(1-x) \frac{\Delta_\perp}{2},
    \textbf{k}'_\perp=\textbf{ k}_\perp +(1-x) \frac{\Delta_\perp}{2},
\end{eqnarray}
where $t=\sqrt{-\Delta_\perp}$. The electromagnetic form factors $F_M(t)$ and the gravitational form factors $A_M(t)$ are expressed in the form of GPD as \cite{Chavez:2021llq}
\begin{eqnarray}
    F_M(t)= \int_{0}^{1} dx  H(x,0,t), \quad
    A_M(t)= \int_{0}^{1} dx x H(x,0,t).
    \end{eqnarray}

\begin{figure}
    \centering
    (a){\label{4figs-e} \includegraphics[width=0.45\textwidth]{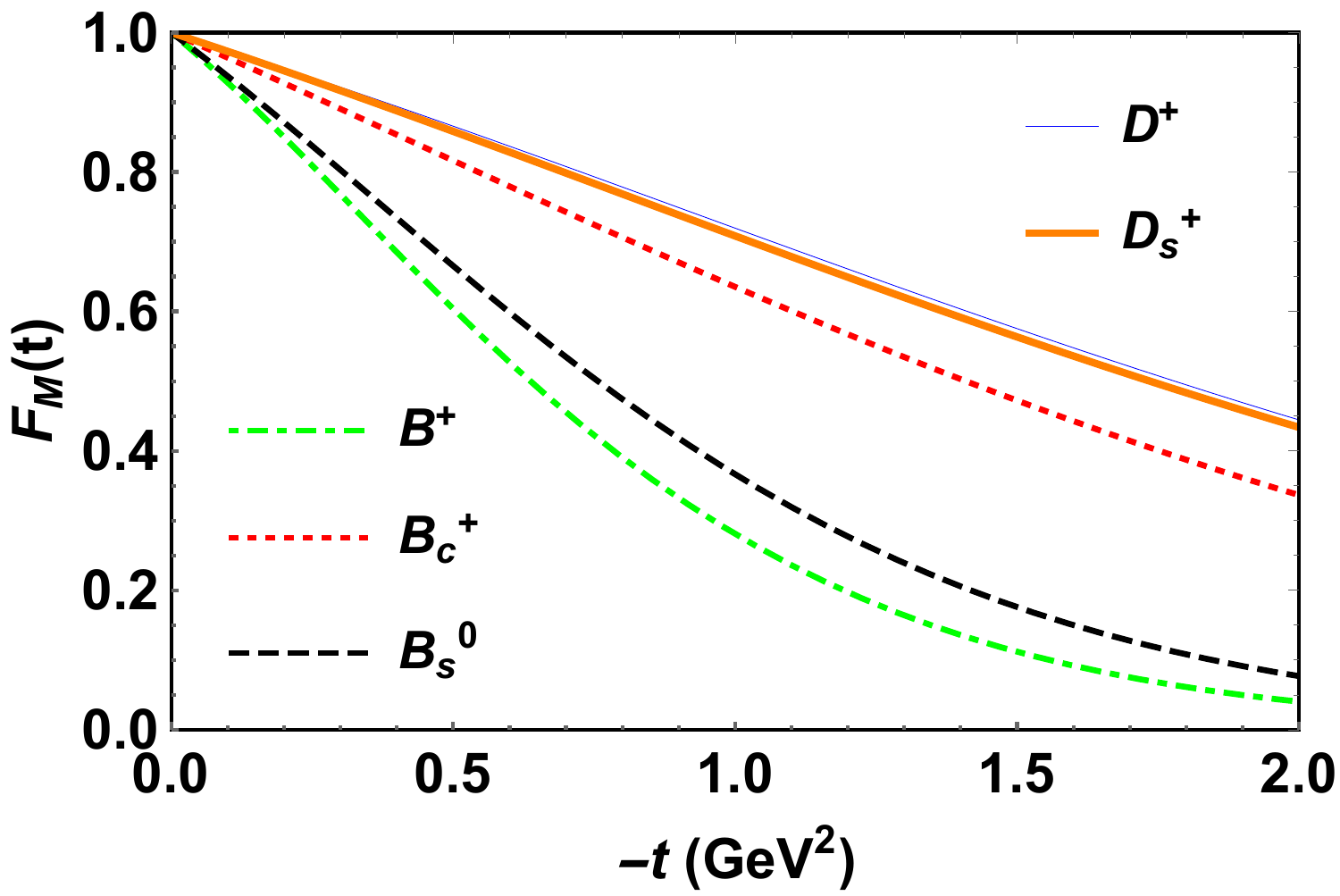}}%
(b){\label{4figs-f4} \includegraphics[width=0.45\textwidth]{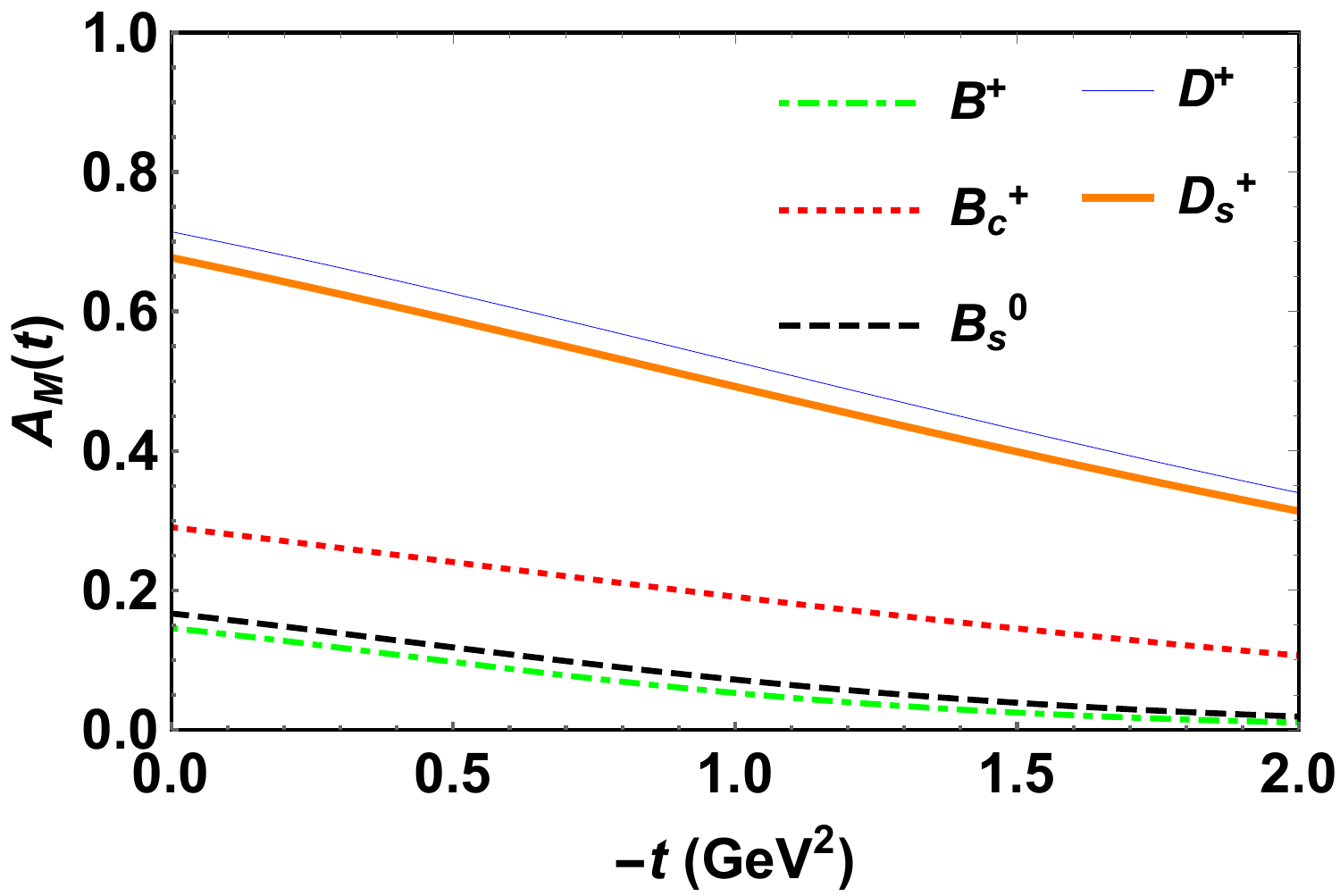}}%
    \caption{Comparison of electromagnetic and gravitational form factors for $B$- and $D$-mesons}.
    \label{a1a1}
\end{figure}
\section{Conclusion}
We have solved the quark-quark correlation function to get the unpolarized GPD and TMD for spin-$0$ heavy mesons. The TMDs provide the information about PDFs, which have been plotted for different particles in Fig. \ref{pdf}. The EMFFs and GFFs are extracted from $H(x,0,t)$ GPD. The EMFFs and GFFs have been plotted with transverse momentum differences $t$ in Fig. \ref{a1a1}.

\section{Acknowledgement} H.D. would like to thank  the Science and Engineering Research Board, Department of Science
and Technology, Government of India through the grant (Ref No.TAR/2021/000157) under TARE
scheme for financial support.

\end{document}